# The small mounds of Bayuda region

**Amelia Carolina Sparavigna**
Dipartimento di Fisica, Politecnico di Torino
Corso Duca degli Abruzzi 24, Torino, Italy

**Abstract**: The Great Bend of the river Nile contains the Bayuda region with its volcanic core. Along the river, a fertile strip of land has attracting human settlement for thousands of years and is then rich of archaeological sites. The distribution of the sites near the Nile can be detected using Google Maps imagery. We can see many area covered by small mounds, probably burial sites. Some of the archaeological places are currently under the water of the Merowe Dam. With the satellite imagery, we have a portrait of the area close the dam before the closing of its gates.

For most of its course, the river Nile flows south to north, but in the heart of the Sahara desert, it turns southwest and flows away from the sea before resuming its northward journey, creating the so-called Great Bend. This turn of the river's flow is due to the tectonic activity of the Nubian Swell, which is a geologic structural uplift in the northern Africa that trends east-west [1]. This structure is separating the Nile of Egypt from the Sudanese basin of the river. The Nubian Swell has some of its parts that are still active. The river passes the uplift through geologic fractures and faults: four of the six Cataracts of the Nile occur during this passage.

The first part of the Great Bend of the river, approximately from the Fourth to the Sixth Cataract, is confining the desert region of Bayuda. This region has a volcanic core, where more than 90 eruptive centres were constructed over Precambrian and Paleozoic granitic rocks. It is interesting to note that the analysis of the youngest basalts of the Bayuda volcanic field are post-dated with respect to the last period of moist climate in Sudan, period that ended as recently as about 5000 years ago [2]. One of the young lava flows at Bayuda was dated at about 1100 before present [3]. The satellite imagery shows quite clearly the volcanic field. Moreover, with a proper image processing enhancement, it is possible to observe all the crater-like structures displayed by the region [4,5].

Other features of the Nile in these Sudanese regions are the Cataracts. There are parts where the river is full of boulders protruding from its bed: for this dangerous local character, an unobstructed sailing of boats from the Equatorial Africa to Egypt was impossible. Moreover, the floodplains turn out to be narrow or nonexistent, strongly constraining an agricultural development. These two problems on navigation and agriculture are important facts to consider in understanding why this region is not so populated in comparison with other regions close the Nile. The First Cataract at Aswan marks the boundary between the Egyptian Nile to the north and the less fertile Nile immediately to the south [1]. This is the traditional boundary between Egypt and the interior of Africa, and historically the border between Egypt and Nubia.

In this desert volcanic region, the Nile Valley has been attracting human settlement for thousands of years. Now, close to the Fourth Cataract, where the river splits into multiple branches with islands in between, the Merowe Dam is impeding the free flow of waters. The main purpose for building this dam was the generation of electricity. A significant portion of the land between the Fourth and Fifth Cataracts has been inundated by the reservoir lake [6]. This area has been densely populated through nearly all periods of prehistory and history, but, as told in Wikipedia, very little archaeological work has ever been conducted in this particular region [7]. The item is continuing, telling that "surveys have confirmed the richness and diversity of traceable remains, from the Stone Age to the Islamic period" and that several institutions have been involved in salvage archaeology in this region (see for instance Refs.[8,9]). According to Wikipedia, the "main problems are the shortness of the remaining time and limited funding. Unlike the large UNESCO campaign conducted in Egypt before the completion of the Aswan High Dam, when more than a thousand archaeological sites … were moved to prevent them from drowning …, work at the Fourth Cataract is much more restricted".

In Figure 1, it is shown in the upper-left panel, the Merowe Dam under construction, obtained from Google Maps. The upper images are antecedent that given in Ref.6 and reproduced in the lower panel, which is showing the region covered by water after the closing of the gates.

A detail of the area before inundation is shown in the upper-right part of the figure. You can see the ground covered by small dots that seem mounds, round with flat top or with a depression or hole in the middle as small volcanic cones, and by circles with smaller mounds inside. Have these mounds a natural origin as the Mima mounds, that are those natural mounds found in the north-western America? [10] Or, had these mounds a human origin in modern or prehistoric times?

First of all, let us check the presence of other such places near the Nile River. We can find more that forty places following the bend of the Nile from Merowe Dam till Gananita, of course, counting only those where the resolution of the maps is high enough to allow a proper observation. We can admit that a larger number surely exists. In Fig.2 and 3, we can see some mounds in the sites at the coordinates reported in the images: sometimes a clear triangular tail is visible, as a pathway to the mound. Of course, the author cannot tell the age of them.

As observed in Refs.11-13, mounds and stone circles are present in Neolithic burial sites of Arabia. Among the Neolithic structures in Arabia there are also lines and "desert kites", which were kite-shaped stone fences, probably used as animal traps. Figure 4 shows for comparison some of the keyhole shaped mounds at Khaybar, Arabia, a place where thousands of tumuli and stone fences, keyhole shaped, kite shaped and circular, cover extensive areas [14]. Who is writing is supposing that objects shown in Figures 1-3 belong to burial places, but a further discussion from an archaeological point of view of these Bayuda sites is impossible. Moreover, the author is not able to give specific references. In the public domain the author found two pages discussing mounds and keyhole tombs [15,16]. In particular, [16] is telling that in Ahaggar, a quite far region indeed, there are "passage tombs and enclosures" or "keyhole tombs" which leads to the corridor to the east, the central mound is sometimes a crater. This seems a description quite suitable for Figures 1-3, and also 4. In Figure 5, two curious structures with many tails are shows.

Here again we have used the Google imagery with a proper image processing enhancement to have a survey of the Bayuda region. There are many places that can be observed, which seem to be archaeological sites. For the region, which is currently occupied by the basin of the Merowe Dam, the Google Maps are free evidences of what was the archaeological site distribution before inundation.

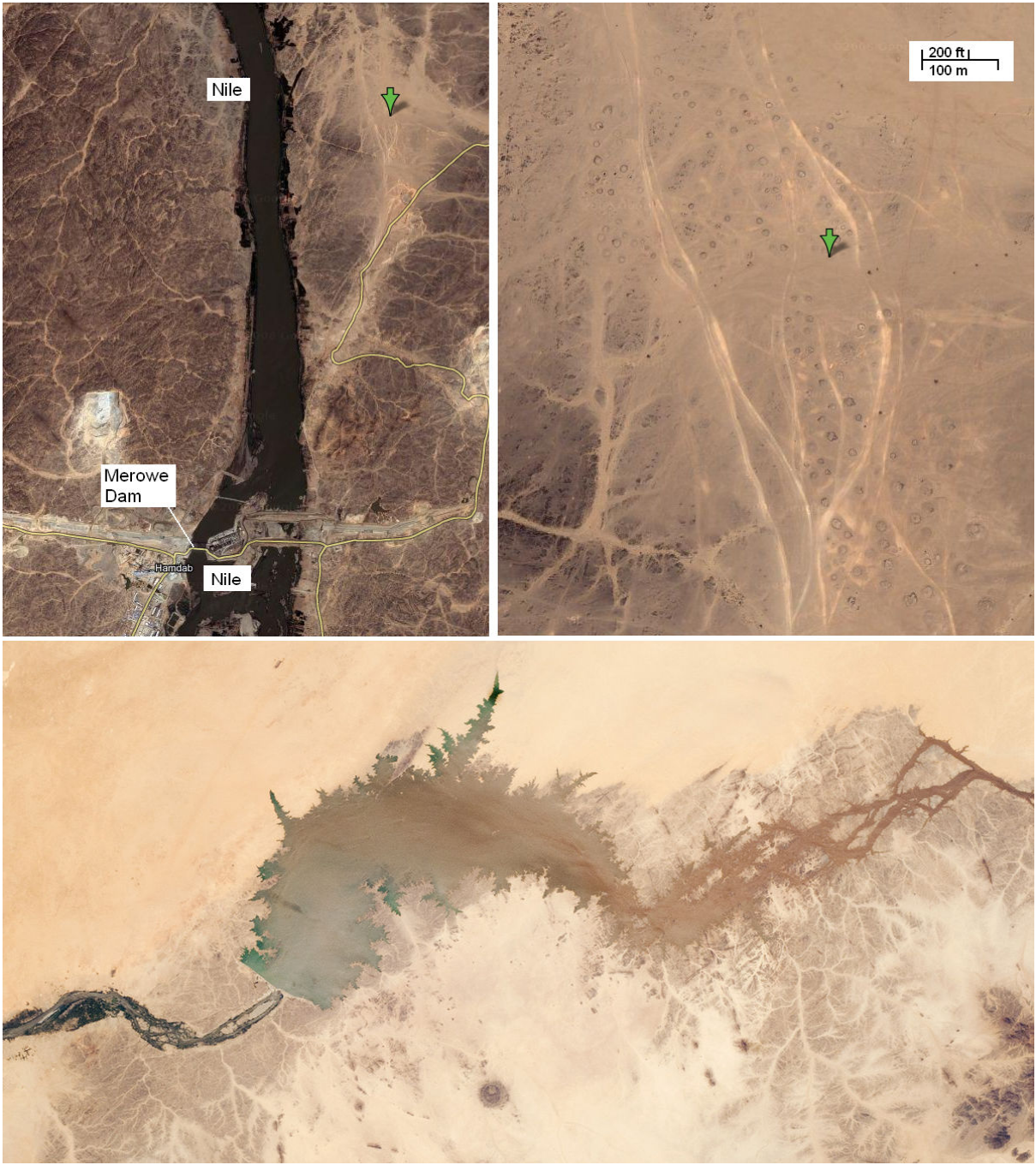

Figure 1: On the upper-left panel, an image adapted from an original Google Maps one, where it is possible to see the Merowe Dam under construction. The images displayed by Google Maps are antecedent that given in Ref.6 and reproduced in the lower panel (NASA's Earth Observatory), with the region covered by water. A detail of the area is shown in the upper-right part of the figure. You can see small dots that could be small mounds, and stone circles with dots inside.

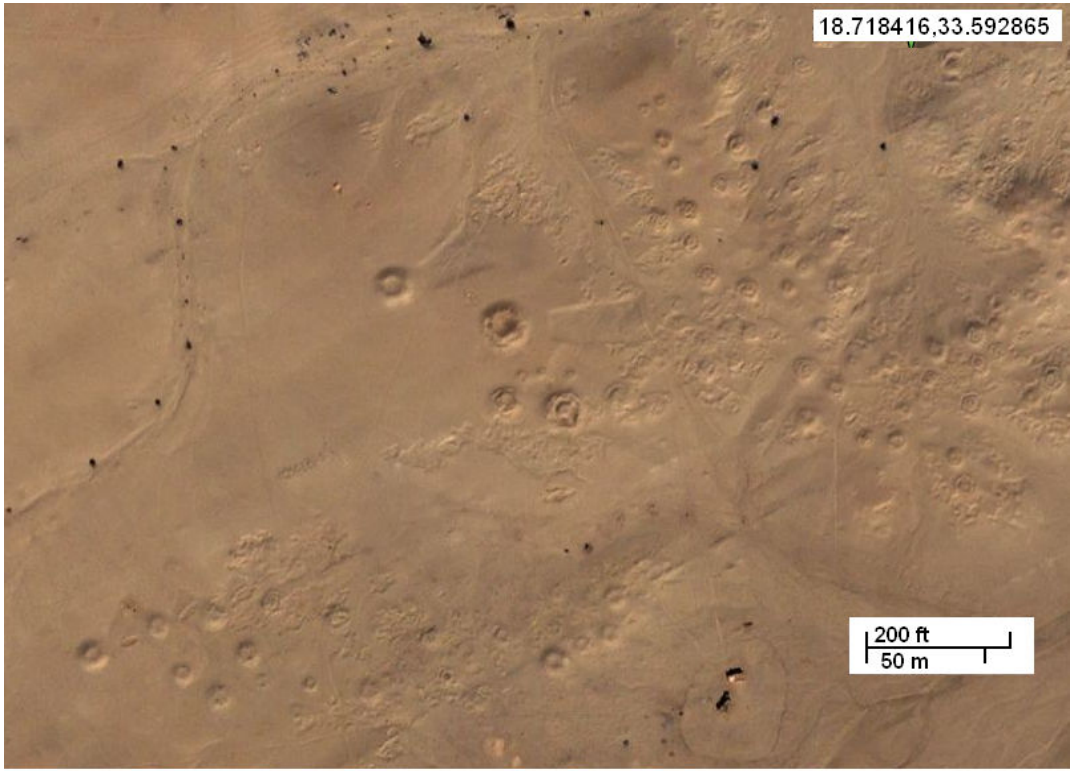
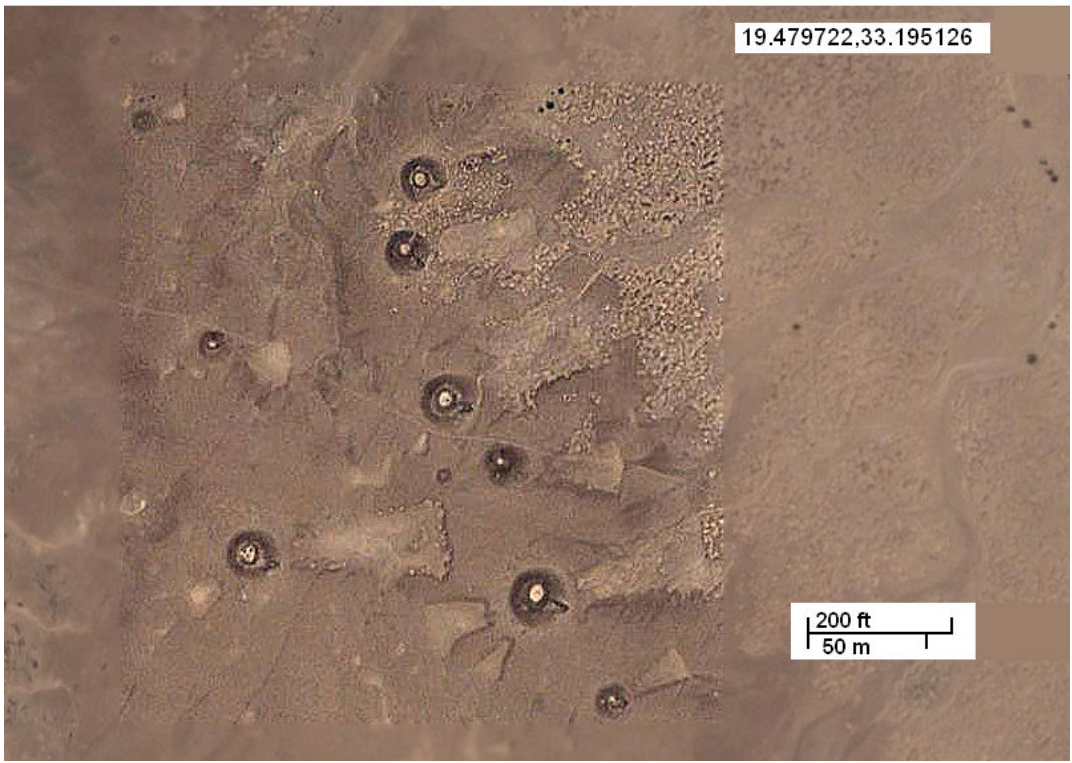

Figure 2: These two images show two areas with some objects that seem mounds or stone circles with tails, or pathways, with a predominant east orientation. The images have been obtained adapting the Google Maps. The corresponding coordinates and scales are shown in the images. The lower image was prepared using a wavelet filter to enhance the structure of the tails.

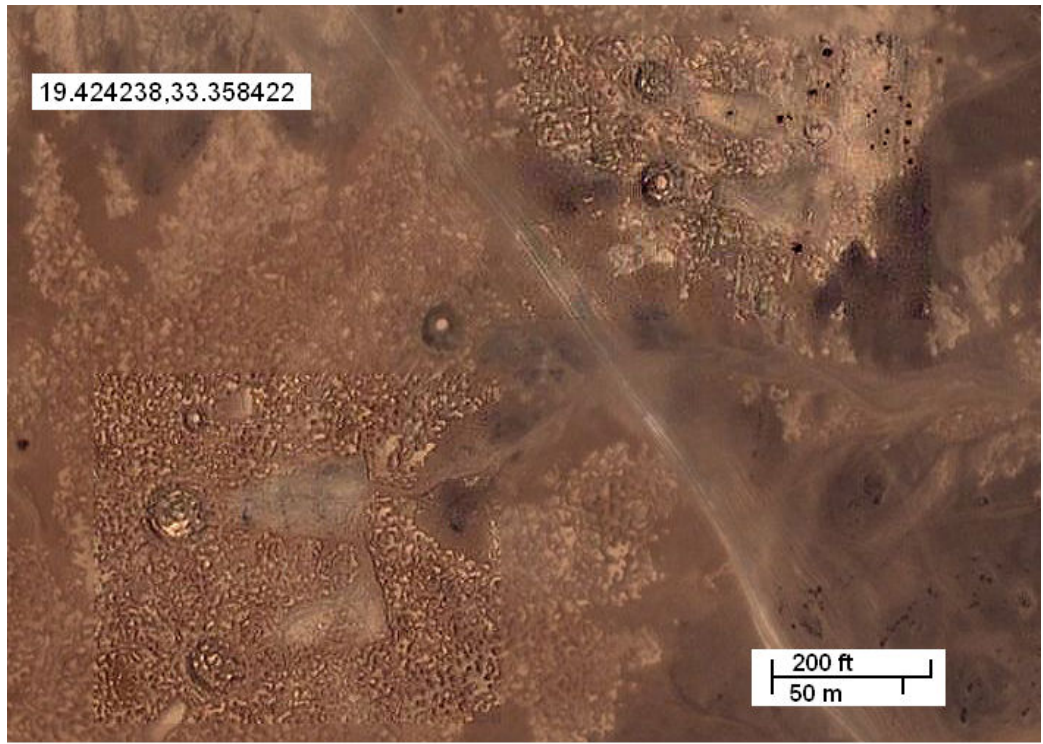

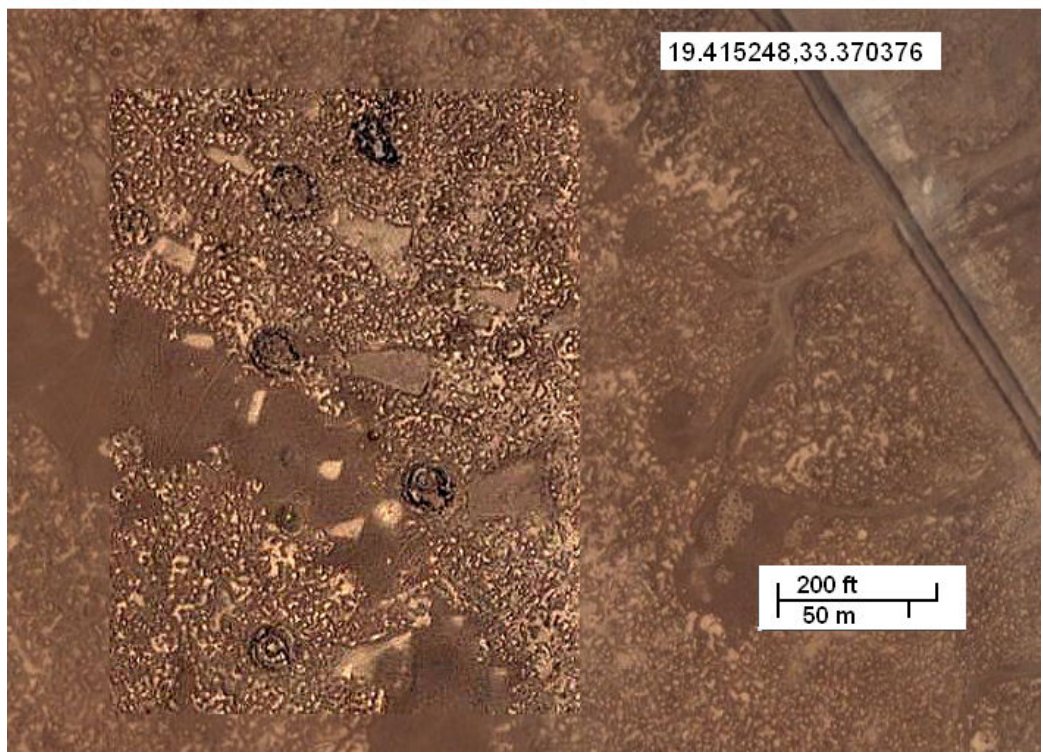

Figure 3: Other two areas with some mounds that have a keyhole shape. The images have been obtained adapting the Google Maps. The corresponding coordinates and scales are shown in the images. The part of the image containing the mounds has been enhanced with image processing.

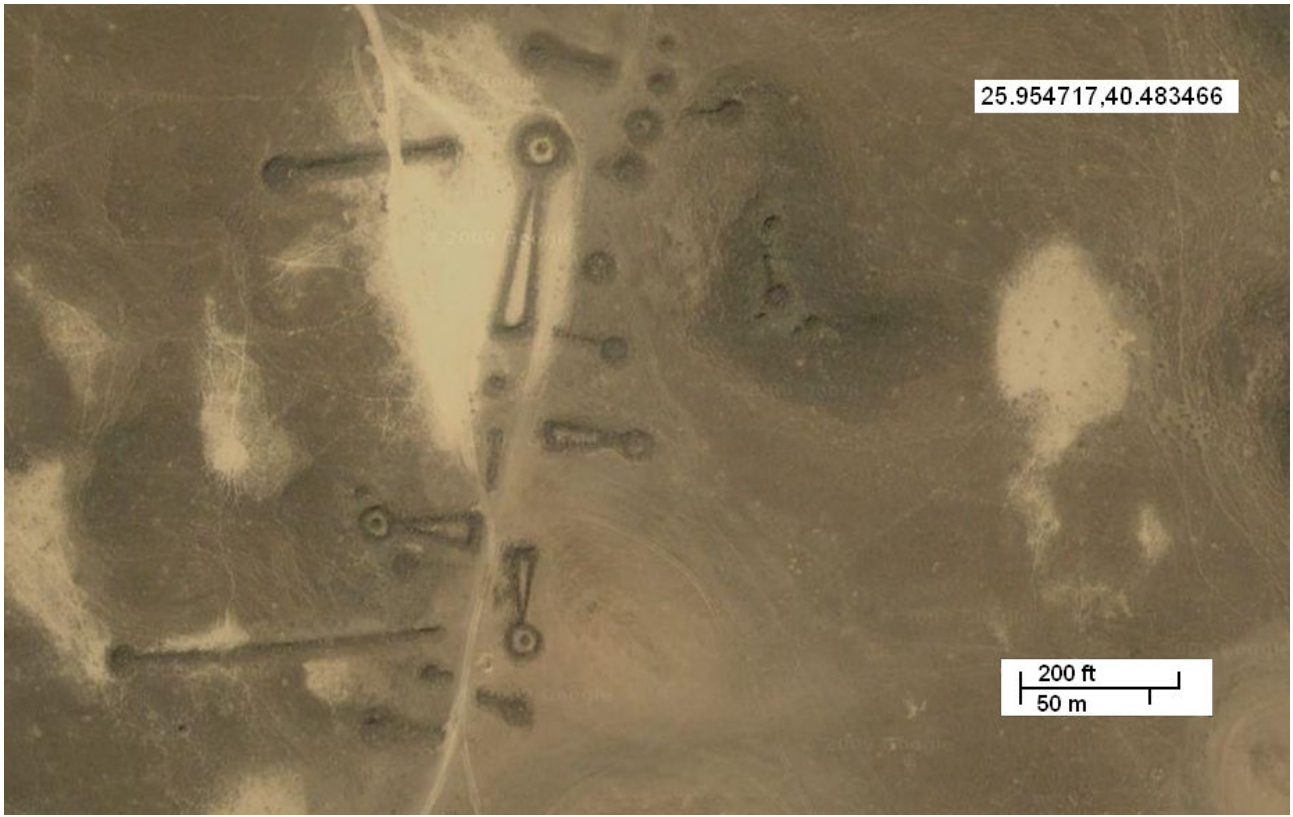

Figure 4: The mounds with keyhole shapes at Khaybar in Arabia (adapted from the Google Maps)

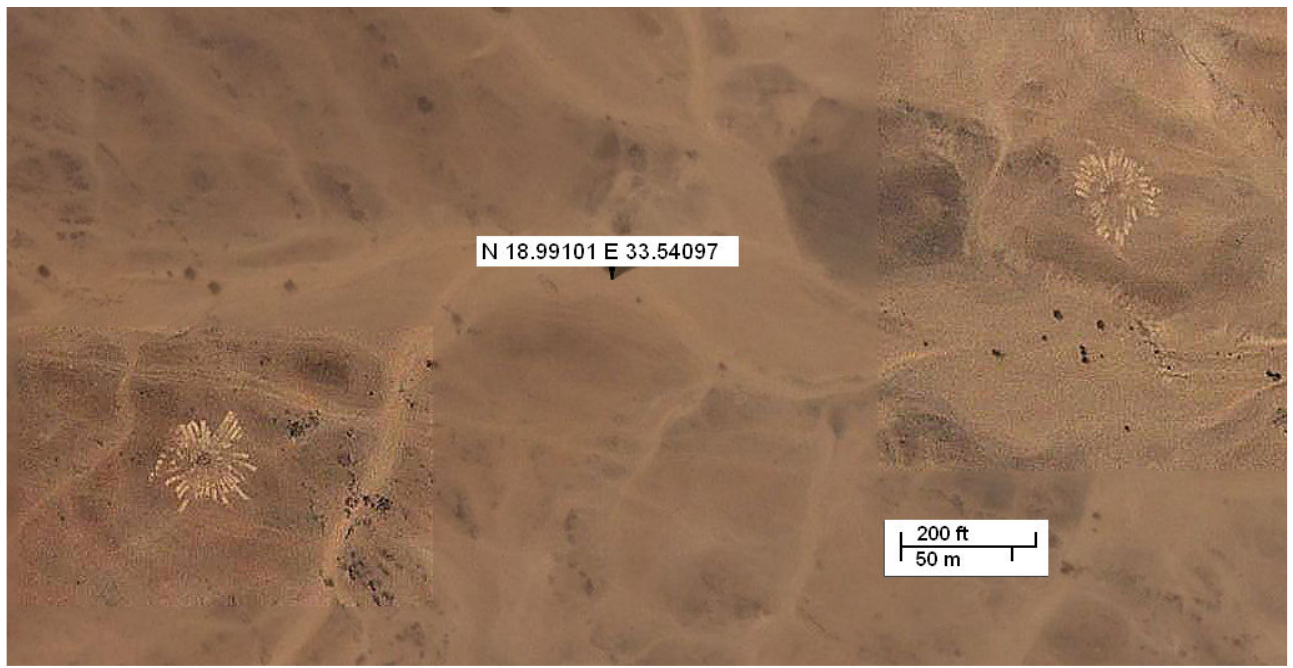

Figure 5: Two curious star-like structures in Bayuda (adapted from the Google Maps)